\documentclass[%
 reprint,
 amsmath,amssymb,
 aps,
]{revtex4-2}

\usepackage[pdftex]{graphicx}% Include figure files
\usepackage{dcolumn}% Align table columns on decimal point
\usepackage{bm}% bold math
\usepackage{array}
\usepackage{booktabs}
\usepackage{multirow}

\usepackage{ulem}
\usepackage{color}

\usepackage{pdfpages}
\usepackage{hyperref}% add hypertext capabilities
\makeatletter
\AtBeginDocument{\let\LS@rot\@undefined}
\makeatother
\begin{document}

\preprint{APS/123-QED}

\title{Possible Realization of Kitaev Spin Liquids in van der Waals Heterostructures of $\alpha$-RuCl$_3$ and Cr$X_3$ ($X$=Cl and I)}% Force line breaks with \\

\author{Lingzhi Zhang}
\author{Yukitoshi Motome}%
% \email{motome@ap.t.u-tokyo.ac.jp}
\affiliation{%
Department of Applied Physics, The University of Tokyo, Bunkyo, Tokyo 113-8656, Japan
}%

\begin{abstract}
  Despite the presence of the exact solution and well-established recipe for its realization, the quest for 
  the Kitaev spin liquid in real materials remains exceptionally challenging. 
  Among many magnets, $\alpha$-RuCl$_3$ emerges as a prime candidate, 
  albeit the hallmarks of the spin liquid manifest only within the specific region 
  where the zigzag-type antiferromagnetic order is suppressed by an applied magnetic field. 
  Here, we propose the possible realization of the Kitaev spin liquid at zero field
  by making van der Waals heterostructures of $\alpha$-RuCl$_3$ and a ferromagnet Cr$X_3$ ($X$=Cl and I). 
  Using {\it ab initio} calculations, we find that in the case of $X$=Cl  
  the zigzag order is suppressed by the proximity effect of the ferromagnetic CrCl$_3$ layer, 
  while the Kitaev interaction is still relevant in the $\alpha$-RuCl$_3$ layer. 
  Notably, the induced Ru moment is close to the value observed in the spin liquid region of the bulk material,
  signifying the possibility of the Kitaev spin liquid at zero field.
  In contrast, in the case of $X$=I, 
  the system is on the verge of an insulator-metal transition by carrier doping through interlayer hybridization. 
  Our results indicate that van der Waals heterostructures provide a platform for studying not only
  the magnetic but also the electronic properties of Kitaev magnets.
  \end{abstract}
  
  \maketitle
  
  %\tableofcontents
  {\it Introduction}.---
  Since the successful exfoliation of graphene~\cite{doi:10.1126/science.1102896}, 
  van der Waals (vdW) materials have played a pivotal role in advancing 
  two-dimensional physics~\cite{geim2007rise, doi:10.1126/science.1158877, wang2012electronics, doi:10.1021/cr300263a, doi:10.1126/science.aac9439, Park_2016, manzeli20172d, burch2018magnetism}.
  The research forefront has been explosively expanded to a diverse range of materials, such as elemental analogues of graphene~\cite{PhysRevB.50.14916, PhysRevLett.102.236804, PhysRevLett.108.155501, PhysRevLett.108.245501, Davila_2014, zhu2015epitaxial},  
  transition metal dichalcogenides~\cite{wang2012electronics, splendiani2010emerging, PhysRevLett.105.136805, 
  doi:10.1126/science.1250140, tong2016concepts}, 
  and atomically thin magnets~\cite{lee2016ising, gong2017discovery, burch2018magnetism, huang2017layer, doi:10.1126/science.aav4450}.
  Furthermore, the weak interlayer vdW force offers a wide platform for making heterostructures with emergent properties that cannot be found in %independent 
  individual materials~\cite{geim2013van, liu2016van, gibertini2019magnetic, li2020general}.
  Moir\'e superstructures in twisted multilayers have opened up yet another exciting avenue for manipulating electron correlation effects, 
  exemplified by twisted bilayer graphene which realizes the Mott insulating state and superconductivity~\cite{cao2018unconventional, andrei2020graphene}.

  Among the vdW materials, 
  a transition metal trihalide $\alpha$-RuCl$_3$ has garnered significant attention 
  due to its potential to realize the quantum spin liquid predicted in the Kitaev model~\cite{PhysRevB.90.041112}.
  The model is defined on a honeycomb lattice with bond-dependent Ising-type interaction between spin-$1/2$ moments, and the ground state is exactly obtained as a quantum spin liquid %where the spin degree of freedom is fractionalized into 
  with fractional excitations, itinerant Majorana fermions and localized $Z_2$ fluxes, which can be utilized for topological quantum computation~\cite{KITAEV20062}. 
  Since it was pointed out that spin-orbit entangled moments can yield the Kitaev-type interaction~\cite{PhysRevLett.102.017205}, 
  many materials with both electron correlation and spin-orbit coupling, called the spin-orbit coupled Mott insulators, 
  have been investigated as potential candidates for the 
  Kitaev spin liquid (KSL)~\cite{Winter_2017, takagi2019concept, doi:10.7566/JPSJ.89.012002, Motome_2020, TREBST20221}.
  Notably, $\alpha$-RuCl$_3$ is a primary candidate, 
  as it exhibits hallmarks of the KSL, 
  as identified by several experimental techniques, such as
  Raman scattering~\cite{PhysRevLett.114.147201, nasu2016fermionic, wang2020range},
  inelastic neutron scattering~\cite{banerjee2016proximate, do2017majorana, banerjee2018excitations},
  and half-quantization of the thermal Hall conductivity~\cite{kasahara2018majorana}
  despite remaining controversies~\cite{PhysRevB.102.220404, doi:10.1126/science.aay5551, czajka2021oscillations, bruin2022robustness, PhysRevX.12.021025, czajka2023planar}.

  Despite the promising hallmarks, the problem is that the KSL appears to 
  manifest only under the magnetic field at low temperatures.
  This issue stems from the fact that, in the absence of the magnetic field, the system exhibits a zigzag-type magnetic order 
  due to parasitic magnetic interactions beyond the Kitaev model~\cite{PhysRevB.91.144420, PhysRevB.92.235119, PhysRevB.93.134423}. 
  The spin liquid nature is observed only in a narrow region where the zigzag order is 
  destroyed by an applied magnetic field of $\sim 8$~T.
  Extensive efforts have been made to suppress the zigzag order by mitigating the parasitic interactions. 
  For instance,
  theoretical research has suggested that lattice expansion can suppress the non-Kitaev interactions~\cite{PhysRevB.93.214431, PhysRevB.98.121107, PhysRevB.103.L140402}, 
  stimulating the fabrication of thin films and their heterostructures~\cite{ziatdinov2016atomic, doi:10.1021/acs.nanolett.6b00701, ZHOU2019291, doi:10.1021/acs.nanolett.9b01691, PhysRevB.100.165426, wang2022direct}.
  Nonetheless, establishing a method to realize the KSL at zero field remains a challenging issue, 
  despite its crucial significance for experimental accessibility.
  This inability to access the KSL in wider conditions hampers in-depth experimental investigations,
  which also highlights the significance of experimental attempts from different aspects in different platforms.

  In this Letter, 
  we propose vdW heterostructures of $\alpha$-RuCl$_3$ with other vdW magnets to explore such platforms for KSL.
  The key idea is to suppress the zigzag order through the proximity effect of vdW magnets.
  Despite the rapid development of thin films of vdW magnets, no attempt was made to fabricate heterostructures including $\alpha$-RuCl$_3$ to our knowledge. 
  Here, we focus on Cr$X_3$ with $X$=Cl and I as the counter compound, 
  which were recently shown, respectively, 
  to exhibit in-plane and out-of-plane ferromagnetism in their monolayer form~\cite{huang2017layer,doi:10.1126/science.abd5146}. 
  Based on {\it ab initio} calculations, we show that an internal magnetic field and interlayer hybridization from the Cr$X_3$ layer %affects 
  have a profound impact on the magnetic and electronic properties of the $\alpha$-RuCl$_3$ layer, 
  offering a promising route to a possible KSL at zero field.

\begin{figure}[t]
    \centering
    \includegraphics[width=1.0\columnwidth]{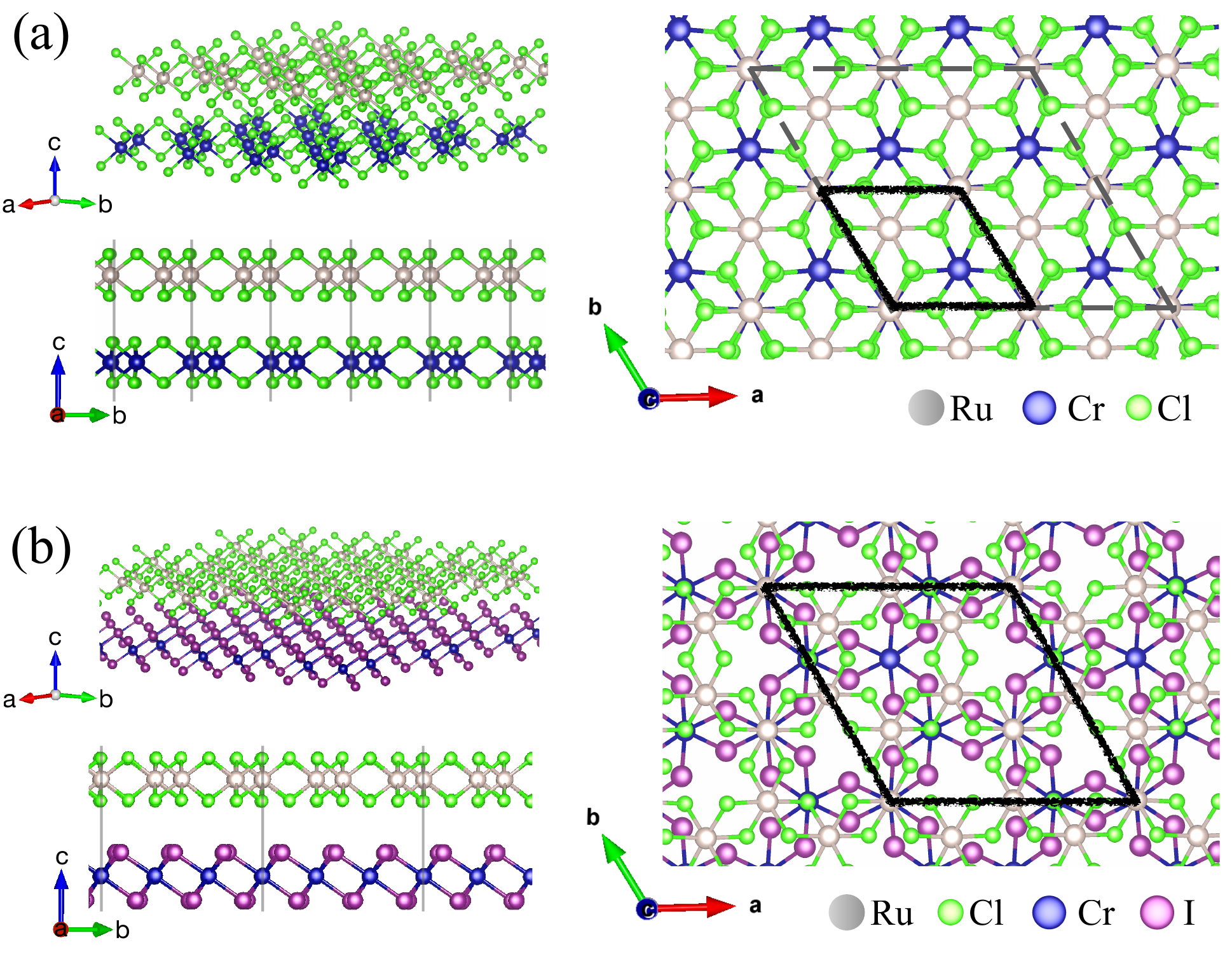}
  
    \centering
    \caption{
  Structures of the heterostructures of $\alpha$-RuCl$_3$ and Cr$X$$_3$ ($X$=Cl and I).
  (a) A bird's-eye view (upper left), side view (lower left), and top view (right) of $\alpha$-RuCl$_3$/CrCl$_3$.
  (b) Corresponding pictures for $\alpha$-RuCl$_3$/CrI$_3$. 
  The gray, blue, green, and purple spheres represent Ru, Cr, Cl, and I atoms, respectively. 
  The black rhombus denotes the unit cell, while the gray dashed rhombus in (a) is a $2\times 2$ supercell used in the {\it ab initio} calculations.
  }
  \label{fig:structure}
  \end{figure}

  {\it Methods}.---
  We consider two heterostructures: 
  One consists of an $\alpha$-RuCl$_3$ layer and a CrCl$_3$ layer ($\alpha$-RuCl$_3$/CrCl$_3$) [Fig.~\ref{fig:structure}(a)], and the other
  consists of an $\alpha$-RuCl$_3$ layer and a CrI$_3$ layer ($\alpha$-RuCl$_3$/CrI$_3$) [Fig.~\ref{fig:structure}(b)]. 
  For both cases, the lattice structure of each layer is taken from the bulk materials with $C2/m$ symmetry~\cite{PhysRevB.92.235119, 10.1063/1.1725428, doi:10.1021/cm504242t}. 
  For $\alpha$-RuCl$_3$/CrCl$_3$, because of the small lattice mismatch between $\alpha$-RuCl$_3$ ($5.97$~\r A) and CrCl$_3$ ($5.96$~\r A),
  we prepare the heterostructure supercell by stacking each unit cell as is, which contains $16$ atoms in total. 
  Meanwhile, for $\alpha$-RuCl$_3$/CrI$_3$, due to the large lattice mismatch between $\alpha$-RuCl$_3$  ($5.97$~\r A) and CrI$_3$ (6.86 \r A),
  we prepare the heterostructure supercell by $2 \times 2$ $\alpha$-RuCl$_3$ %
  and $\sqrt{3}\times\sqrt{3}$ CrI$_3$, containing $56$ atoms. 
  We optimize the lattice structures via the nonrelativistic {\it ab initio} calculations described below, 
  by relaxing the atomic positions with fixing the in-plane lattice constants of $\alpha$-RuCl$_3$ at the bulk value. 
  After the relaxation, the interlayer distance is optimized to $3.01$~\r A for $\alpha$-RuCl$_3$/CrCl$_3$ and $3.22$~\r A for $\alpha$-RuCl$_3$/CrI$_3$.
  
  Considering the wide applicability of density functional theory (DFT) and the consistency between previous DFT calculations and experiments for $\alpha$-RuCl$_3$
  ~\cite{PhysRevB.93.214431, PhysRevB.93.155143, PhysRevLett.118.107203, winter2017breakdown, PhysRevB.96.054410},
  we perform DFT calculations for our systems
  using the OpenMX code~\cite{PhysRevB.67.155108, PhysRevB.69.195113}.
  We use the Perdew-Burke-Ernzerhof generalized gradient approximation (GGA-PBE) functional~\cite{PhysRevLett.77.3865}
  including $U$ corrections for Ru $d$ orbitals and Cr $d$ orbitals, as implemented in the GGA+$U$ method~\cite{PhysRevB.57.1505};
  we take $U=1.5$~eV for Ru and $U=2.0$~eV for Cr~\cite{PhysRevB.97.220401,Lado_2017}. 
  The kinetic energy cutoffs are set to be $1500$~Ry. 
  VdW corrections are also included through the DFT-D2 scheme of Grimme~\cite{https://doi.org/10.1002/jcc.20495}.
  In the self-consistent calculation of the electron density,
  we take $16 \times 16 \times 1$ $\bm{k}$-point mesh for $\alpha$-RuCl$_3$/CrCl$_3$ and
  $10 \times 10 \times 1$ $\bm{k}$-point mesh for $\alpha$-RuCl$_3$/CrI$_3$.
  We construct maximally-localized Wannier functions~\cite{PhysRevB.56.12847, PhysRevB.65.035109} 
  for the electronic band structures obtained by the DFT calculation 
  and use them to compute the partial density of states (PDOS) and effective exchange constants.
  For the latter, we perform the second-order perturbation
  for the multiorbital Hubbard model for $t_\text{2g}$ orbitals of Ru atoms, following the previous studies~\cite{PhysRevB.93.214431, PhysRevLett.113.107201, PhysRevB.101.100410}.
  In this model, we set the effective onsite Coulomb interaction $\tilde{U}$, the Hund's-rule coupling $J_{\text{H}}$, and the spin-orbit coupling coefficient $\lambda$ to $3.0$~eV, $0.6$~eV, and $0.15$~eV, respectively~\cite{PhysRevB.93.214431}.

  \begin{table}[]
    \centering
    \caption{
  Energy per Ru atom and magnetic moments at Ru and Cr atoms for different magnetic states in $\alpha$-RuCl$_3$/CrCl$_3$ obtained by the {\it ab initio} calculations:
  ``in" and ``out" mean that the magnetic moment is along the in-plane and out-of-plane direction, respectively,
  and ``FM", ``N\'eel", and ``zigzag" denote ferromagnetic, N\'eel, and zigzag orders, respectively.
  The prime indicates that the Cr moments are antiparallel to the Ru ones (otherwise, they are parallel). 
  The energies are measured from the most stable state where the in-plane FM order in each layer is coupled antiferromagnetically.
  }
  
    \begin{ruledtabular}
    \begin{tabular}{ccclcc}
    \multicolumn{2}{c}{magnetic state}                      & \multicolumn{2}{c}{\multirow{2}{*}{energy/Ru (meV)}}                                 & \multicolumn{2}{c}{moment ($\mu_{\rm B}$)} \\
    Ru                         & Cr                         & \multicolumn{2}{c}{}                       & Ru                   & Cr                 \\ \hline
    in-FM                      & in-FM'                     & \multicolumn{2}{c}{0.000}                  & 0.703                & 3.28               \\
    in-FM                      & out-FM                     & \multicolumn{2}{c}{0.365}                  & 0.703                & 3.28               \\
    in-FM                      & in-FM                      & \multicolumn{2}{c}{0.775}                  & 0.704                & 3.28               \\ %\hline
    in-zigzag                  & in-FM                      & \multicolumn{2}{c}{6.32}                   & 0.579                & 3.28               \\
    in-N\'eel                  & in-FM                      & \multicolumn{2}{c}{16.2}                   & 0.293                & 3.28               \\
    out-FM                     & out-FM'                    & \multicolumn{2}{c}{104}                    & 0.832                & 3.28               \\
    out-FM                     & out-FM                     & \multicolumn{2}{c}{105}                    & 0.834                & 3.28               %\\ \hline
    \end{tabular}
    \end{ruledtabular}
    \label{tab:rucl3-crcl3-energy}
  \end{table}

  \begin{table}[]
    \centering
    \caption{
  Similar table for $\alpha$-RuCl$_3$/CrI$_3$: 
  ``cant" means a spin canting, and the other notations are common to Table~\ref{tab:rucl3-crcl3-energy}.
  } 
  
    \begin{ruledtabular}
    \begin{tabular}{ccclcc}
    \multicolumn{2}{c}{magnetic state} & \multicolumn{2}{c}{\multirow{2}{*}{energy/Ru (meV)}} & \multicolumn{2}{c}{moment ($\mu_{\rm B}$)} \\
    Ru                         & Cr                                                   & \multicolumn{2}{c}{}                         & Ru                   & Cr               \\ \hline
    in-FM                      & cant-FM                                              & \multicolumn{2}{c}{0.000}                    & 0.675                & 3.73             \\
    in-FM                      & out-FM                                               & \multicolumn{2}{c}{0.359}                    & 0.673                & 3.73             \\
    in-FM                      & in-FM'                                               & \multicolumn{2}{c}{0.443}                    & 0.678                & 3.73             \\
    in-FM                      & in-FM                                                & \multicolumn{2}{c}{3.14}                     & 0.673                & 3.73             \\ %\hline
    in-zigzag                  & out-FM                                               & \multicolumn{2}{c}{8.42}                     & 0.571                & 3.73               
    \end{tabular}
    \end{ruledtabular}
    \label{tab:rucl3-cri3-energy}
  \end{table}
  
  {\it Results}.---
  For the optimized heterostructures, we investigate the energies of different magnetically ordered states by utilizing 
  the constraint DFT calculation on the direction of magnetic moments.
  The results for $\alpha$-RuCl$_3$/CrCl$_3$ are listed in Table~\ref{tab:rucl3-crcl3-energy}.
  We obtain seven stable magnetic states, and among them the most stable %one 
  is the intralayer-ferromagnetic (FM) and interlayer antiferromagnetic (AFM) state with in-plane magnetic moments of 
  $0.703$~$\mu_{\rm B}$ for Ru and $3.28$~$\mu_{\rm B}$ for Cr.
  Notably, the intralayer FM states with different Cr moment directions are obtained at slightly higher energies: 
  the state with out-of-plane Cr moments at $0.365$~meV and in-plane Cr moments parallel to the Ru ones at $0.775$~meV. 
  This suggests that the Cr layer has in-plane ferromagnetism similar to the monolayer~\cite{doi:10.1126/science.abd5146}, 
  but the magnetic anisotropy is small 
  and the energy scale of the interlayer magnetic coupling in this heterostructure is the order of $0.1$~meV 
  (see Supplemental Material~\cite{SM_MAE}).
  \nocite{IDO2024109093, KAWAMURA2017180}
  The next higher-energy state at $6.32$~meV exhibits a zigzag order in the Ru layer. 
  While the stable magnetic state for the $\alpha$-RuCl$_3$ monolayer is still controversial~\cite{PhysRevB.91.241110, PhysRevB.95.045113, doi:10.1021/acs.nanolett.9b02523, PhysRevB.106.155118}, 
  our result indicates that the heterostructure with CrCl$_3$ favors the FM state in the $\alpha$-RuCl$_3$ layer, 
  rather than the zigzag order that is stable in the bulk.
  We will discuss this interesting point later. 
  The other states have much higher energies; 
  in particular, the states with out-of-plane Ru moments are more than $100$~meV higher, 
  indicating that the strong in-plane magnetic anisotropy in the bulk $\alpha$-RuCl$_3$ is retained even in the heterostructure~\cite{doi:10.1021/cm504242t, PhysRevB.91.094422, modic2021scale, yang2023magnetic}. 
  
  The results for $\alpha$-RuCl$_3$/CrI$_3$ are summarized in Table~\ref{tab:rucl3-cri3-energy}.
  In this case also, we find that the most stable state exhibits FM ordering in each layer, 
  whereas the Cr moments are canted about $30$ degrees from the out-of-plane in the opposite direction to the Ru moments. 
  This is understood from the fact that the monolayer CrI$_3$ is an out-of-plane ferromagnet~\cite{huang2017layer}. 
  Nonetheless, similarly to the case of $\alpha$-RuCl$_3$/CrCl$_3$, the energy differences from the states with other Cr moment directions are 
  rather small, while the zigzag order in the Ru layer leads to relatively higher energy (see Supplemental Material~\cite{SM_MAE}).

  \begin{figure}[t]
    \centering
    \includegraphics[width=1.0\columnwidth]{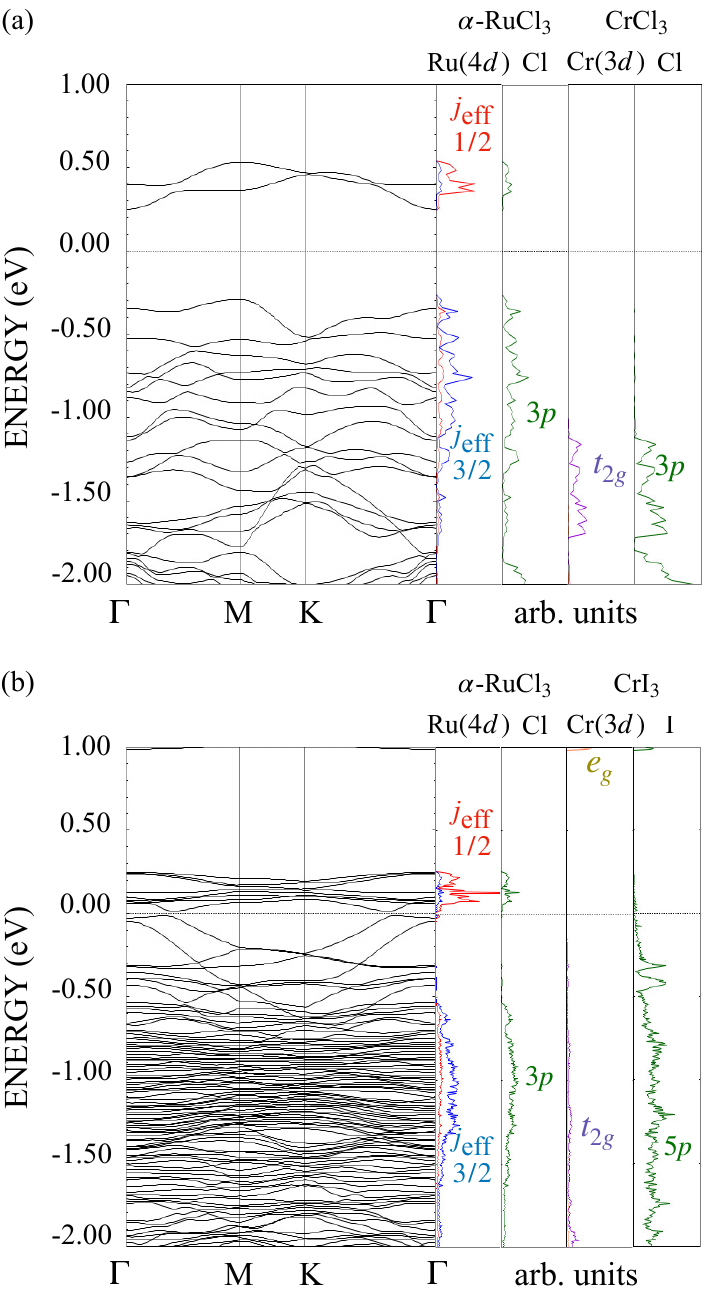}
  
    \centering
    \caption{
  Band structures and PDOS %of 
  in the most stable magnetic states: (a) the Ru in-FM and Cr in-FM' state 
  for $\alpha$-RuCl$_3$/CrCl$_3$ and (b) the Ru in-FM and Cr cant-FM state 
  for $\alpha$-RuCl$_3$/CrI$_3$. 
  The Fermi level is set at zero energy.
  }
  \label{fig:band}
  \end{figure}
  
  For the most stable states obtained above, we calculate the band structures and the PDOS.
  For $\alpha$-RuCl$_3$/CrCl$_3$ [Fig.~\ref{fig:band}(a)], we find that the low-energy state near the Fermi level is dominated by the spin-orbital coupled bands characterized by the pseudospin $j_{\rm eff}=1/2$ and $3/2$ hybridized with $3p$ orbitals of Cl atoms in the $\alpha$-RuCl$_3$ layer. 
  The system is an insulator with the energy gap of $\sim 0.6$~eV opening in the pseudospin bands. 
  The situation is very similar to the bulk
  $\alpha$-RuCl$_3$, where the spin-orbit coupling and electron correlation work cooperatively to 
  realize a spin-orbit coupled Mott insulating state~\cite{PhysRevB.90.041112, sinn2016electronic, PhysRevLett.117.126403}.
  The contribution from CrCl$_3$ appears well below the Fermi level, and does not disturb the low-energy pseudospin bands of Ru atoms.

  \begin{figure}[t]
    \centering
    \includegraphics[width=1.0\columnwidth]{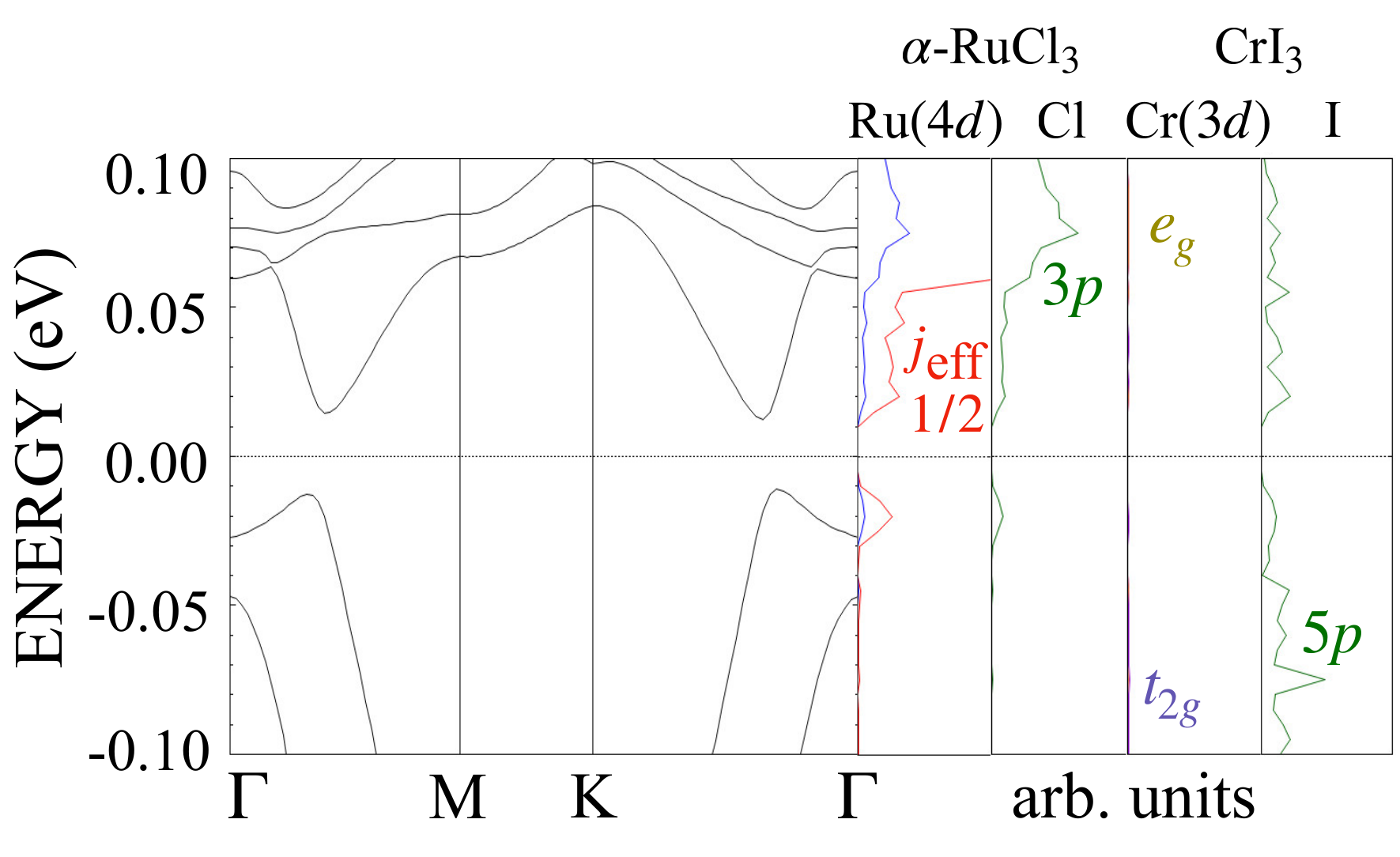}
  \centering
  \caption{
  Enlarged band structure and PDOS near the Fermi level for $\alpha$-RuCl$_3$/CrI$_3$ in Fig.~\ref{fig:band}(b).
  }
  \label{fig:band-rucl3-cri3-d}
  \end{figure}
  
  Meanwhile, in $\alpha$-RuCl$_3$/CrI$_3$, the spin-orbital coupled features are preserved in the pseudospin bands, 
  but the band gap is largely reduced [Fig.~\ref{fig:band}(b)]. 
  This is due to considerable contributions from the $5p$ orbitals of I atoms, 
  in contrast to the Cl $3p$ orbitals in the case of $\alpha$-RuCl$_3$/CrCl$_3$. 
  The enlarged plot near the Fermi level is shown in Fig.~\ref{fig:band-rucl3-cri3-d}.
  We find that there still remains a small gap of $\sim 0.02$~eV, but the system is on the verge of an insulator-metal transition, 
  and the $j_{\rm eff}=1/2$ band is slightly doped below the small gap through the hybridization with the $5p$ bands.

  \begin{figure}[htbp]
    \centering
    \includegraphics[width=1.0\columnwidth]{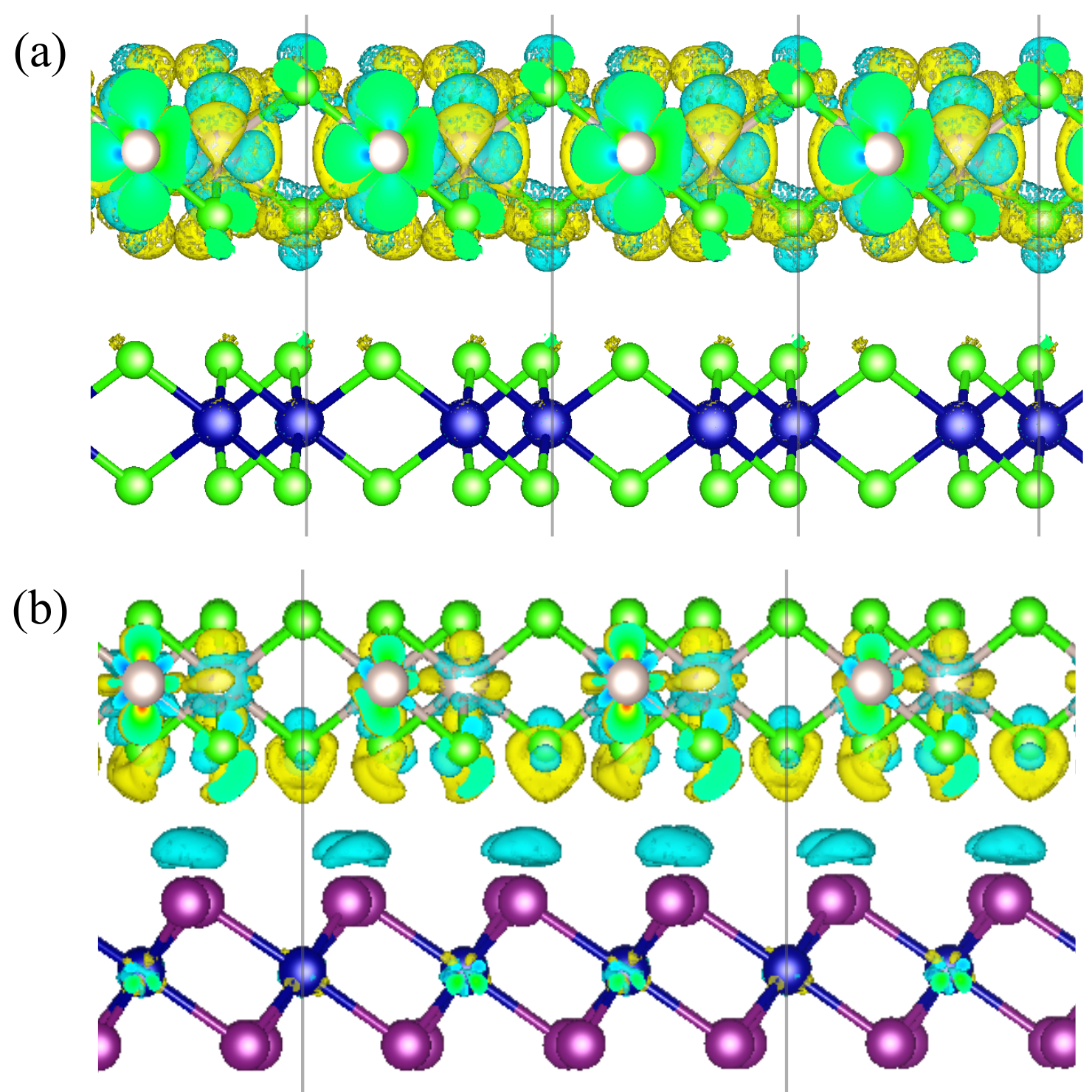}
    \centering
    \caption{
  Charge-density difference $\delta \rho$ plotted with charge isosurface $5 \times 10^{-4}$~$e/$\r A$^3$ of 
  (a) $\alpha$-RuCl$_3$/CrCl$_3$ and (b) $\alpha$-RuCl$_3$/CrI$_3$. 
  Both are viewed from the $a$ axis of the $\alpha$-RuCl$_3$ layer.
  The yellow (cyan) color represents regions with charge accumulation (deficiency) compared to independent monolayers.
  The gray, blue, green, and purple spheres represent Ru, Cr, Cl, and I atoms, respectively.
    }
  
  \label{fig:charge}
  \end{figure}
  
  To understand the differences in the band structures between the two heterostructures,
  we compute real-space distributions of electron charges. 
  Figure~\ref{fig:charge} displays the charge-density difference between the heterostructure and two independent monolayers,
  $\delta \rho({\alpha\text{-RuCl}_3\text{/Cr}X_3})=\rho({\alpha\text{-RuCl}_3\text{/Cr}X_3})-\rho({\alpha\text{-RuCl}_3})-\rho({\text{Cr}X_3}$),
  with the isosurface of $5 \times 10^{-4}$~$e/$\r A$^3$.
  In $\alpha$-RuCl$_3$/CrCl$_3$ [Fig.~\ref{fig:charge}(a)], while the charge distribution in the $\alpha$-RuCl$_3$ layer is modulated,
  the original symmetry is overall retained, with little changes in the interlayer region.
  This is consistent with the robust pseudospin bands in Fig.~\ref{fig:band}(a). 
  In contrast, in $\alpha$-RuCl$_3$/CrI$_3$ [Fig.~\ref{fig:charge}(b)], 
  the charge distributions near the Cl and I atoms at the interface are largely modulated and extended to the interlayer region.
  We observe that approximately $0.021e$ per unit cell of $\alpha$-RuCl$_3$ is transferred from the CrI$_3$ layer. 
  These are consistent with the large modulation of the band structure with carrier doping through the hybridization with $5p$ orbitals of I atoms found in Figs.~\ref{fig:band}(b) and \ref{fig:band-rucl3-cri3-d}.
  
  \begin{table}[]
    \centering
    \caption{
  Effective exchange constants between the $j_{\rm eff}=1/2$ pseudospins at the Ru sites for $\alpha$-RuCl$_3$/CrCl$_3$, 
  in comparison with those for the monolayer and bulk of $\alpha$-RuCl$_3$. 
  The energy unit is meV.
  }
  
    \begin{ruledtabular}
    \begin{tabular}{cccccccc}
                                                                     & $J$    & $K$    & $\Gamma$    & $\Gamma^{\prime}$       &$D$        &$D^{\prime}$    & $K/J$ \\ \hline
        $\alpha$-RuCl$_3$/CrCl$_3$                                   & -0.9   & -5.9   & 6.6         & -1.6                    &-0.1       &-0.1            & 6.6      \\
        $\alpha$-RuCl$_3$ (monolayer)                                & -1.0   & -6.6   & 7.1         & -2.0                    & 0.0       & 0.0            & 6.6      \\
        $\alpha$-RuCl$_3$ (bulk)~\cite{PhysRevB.93.214431}           & -2.2   & -5.0   & 8.0         & -1.0                    & 0.0       & 0.0            & 2.3       \\
        %$\alpha$-RuCl$_3$/CrI$_3$                                    & -0.3   & -7.4   & 5.3         & -1.2                    &-0.9       & 0.2            & 23.8                
    \end{tabular}
    \end{ruledtabular}
    \label{tab:magnetic-constants}
  \end{table}
  
  Given that the spin-orbit coupled Mott insulating nature is preserved in the $\alpha$-RuCl$_3$ layer of $\alpha$-RuCl$_3$/CrCl$_3$, 
  we estimate effective exchange constants between the $j_{\rm eff}=1/2$ pseudospins, using the procedure described above. 
  The general expression of the effective pseudospin %model to describe this system, 
  Hamiltonian for one of three types of bonds on the honeycomb lattice is given by 
  \begin{align}
  {\mathcal H}_{ij}^z=
  {\bf S}_i^{\rm T}\left[ %(
    \begin{array}{ccc}
                            J          
  &  D
  +\Gamma                   
  & -D
  ^{\prime}+\Gamma^{\prime} 
  \\
      -D         
  +\Gamma          
  &              J                   
  &  D
  ^{\prime}+\Gamma^{\prime} 
  \\
       D
  ^{\prime}+\Gamma^{\prime}
  & -D
  ^{\prime}+\Gamma^{\prime}
  &  J
  +K
  \\
    \end{array}
  \right] {\bf S}_j,
  \label{eq:J}
  \end{align}
  where $J$ and $K$ denote the Heisenberg and Kitaev interactions, respectively, 
  $\Gamma$ and $\Gamma^\prime$ denote the symmetric off-diagonal interactions, 
  and $D$ and $D^\prime$ denote the antisymmetric off-diagonal Dzyaloshinskii-Moriya (DM) interactions. 
  The estimates of the exchange constants are listed in Table~\ref{tab:magnetic-constants},
  in comparison with the results for the monolayer (obtained in the present study) and bulk of $\alpha$-RuCl$_3$~\cite{PhysRevB.93.214431}.
  We find that in $\alpha$-RuCl$_3$/CrCl$_3$
  the ratio of $K/J$ is enhanced compared to the bulk, whereas it is comparable to that in the monolayer.
  This indicates that the proximity effect in the heterostructure does not
  disturb the dominant Kitaev physics in the monolayer, despite the charge modulation in Fig.~\ref{fig:charge}(a).
  Thus, the suppression of the zigzag order is ascribed to the synergetic effect 
  between the enhanced dominance of the
  Kitaev interaction from the bulk to monolayer and 
  the internal magnetic field from the ferromagnetic CrCl$_3$ layer in the heterostructure. 
  
  {\it Discussion}.---
  In $\alpha$-RuCl$_3$/CrCl$_3$, the zigzag order in $\alpha$-RuCl$_3$ is suppressed by the proximity effect of CrCl$_3$, 
  and replaced by the FM state with Ru moment of $0.703$~$\mu_{\rm B}$. 
  Notably, this value is close to that in the spin liquid region in the bulk system where the zigzag order is 
  destroyed by an in-plane magnetic field~\cite{PhysRevB.92.235119} and 
  the half-quantization of the thermal Hall conductivity was observed~\cite{kasahara2018majorana, doi:10.1126/science.aay5551}. 
  We confirm that the Ru layer in our heterostructure is indeed in the vicinity of the suppression of the zigzag order, using the exact diagonalization of 24-site clusters (see Supplemental Material~\cite{SM_MAE}).
  This, in conjunction with the dominant Kitaev interaction in Table~\ref{tab:magnetic-constants}, 
  suggests the possibility of KSL at zero field under the proximity effect in the vdW heterostructure.

  Detecting quantum spin liquids in atomically thin films and heterostructures is still a challenge, despite its importance in future applications toward topological quantum computing. 
  Local probes, such as scanning tunneling microscope (STM)~\cite{PhysRevB.102.085412, PhysRevB.102.134423, PhysRevLett.125.227202, PhysRevLett.125.267206, PhysRevLett.126.127201} and atomic force microscopy (AFM)~\cite{PhysRevB.104.085142}, are anticipated as powerful tools to potentially detect and control the Majorana zero mode.
  Besides, Raman scattering spectroscopy is another powerful tool %with its application and achievements in both the bulk and thin film systems
  for thin films as well as bulk~\cite{PhysRevLett.114.147201, nasu2016fermionic, wang2020range, ZHOU2019291, lee2021multiple}.
  It was recently shown that the
  the tunneling spectroscopy~\cite{yang2023magnetic} and
  spin Seebeck effect~~\cite{kato2024spin} 
  can detect Majorana excitations in the KSL.
  Although the above techniques have been developed only recently, our ﬁndings of another platform would stimulate their further advancement

  In $\alpha$-RuCl$_3$/CrI$_3$, 
  while the spin-orbit Mott insulating nature is preserved in the Ru pseudospin bands, 
  the band gap is almost closing and the self-doping occurs via the interlayer hybridization. 
  A similar situation was found in the heterostructure of $\alpha$-RuCl$_3$ and graphene~
  \cite{doi:10.1021/acs.nanolett.9b01691, PhysRevB.100.165426, PhysRevLett.123.237201, PhysRevLett.124.106804}, 
  where the system becomes metallic. 
  Thus, the vdW heterostructures offer a playground for the 
  insulator-metal transition in the spin-orbital coupled systems by carrier doping, 
  which is expected to bring about exotic quantum phenomena such as topological superconductivity~\cite{%YuKitaev2001, Sato_2017, 
  PhysRevB.86.085145, PhysRevB.87.064508, PhysRevB.97.014504, PhysRevB.97.085118}. %, 
  This is complementary to bandwidth control, which has recently been studied for %in 
  the bulk materials Ru$X_3$ ($X$=Cl, Br, and I)~\cite{doi:10.7566/JPSJ.90.123703, PhysRevB.105.L041112, kaib2022electronic, https://doi.org/10.1002/adma.202106831}.

  {\it Summary}.---
  We have investigated the vdW heterostructures composed of the Kitaev candidate $\alpha$-RuCl$_3$ and the ferromagnet Cr$X_3$ ($X$=Cl and I) based on the {\it ab initio} calculations. 
  For $\alpha$-RuCl$_3$/CrCl$_3$,
  we found that the zigzag order in $\alpha$-RuCl$_3$ is replaced by an in-plane FM state 
  with the Ru moment close to that in the region where the hallmarks of KSL were observed in the bulk material. 
  We also showed that the Kitaev interaction between the Ru pseudospins is enhanced from that in the bulk. 
  These suggest that the proximity of CrCl$_3$ promotes the KSL in the $\alpha$-RuCl$_3$ layer even in the absence of the magnetic field.
  In $\alpha$-RuCl$_3$/CrI$_3$, we found that the system is on the verge of 
  insulator-metal transition due to charge transfer from CrI$_3$ via the strong interlayer hybridization, 
  while the spin-orbit coupled Mott insulating nature is preserved. 
  Our findings unveil that the vdW heterostructures
  have the potential to access the KSL physics at zero field, 
  which is hard to achieve in the bulk system, and exotic topological superconductivity.
  Our work opens up possibilities for tailoring the electronic and magnetic properties in the Kitaev magnets by leveraging the diverse range of vdW materials and their flexible stacking control.

  We thank S.-H. Jang, Y. Kato, S. Okumura, and Y.-F. Zhao for fruitful discussions. 
  This research was supported by the JSPS KAKENHI (Nos.~JP19H05825 and JP20H00122) and JST CREST (No.~JP-MJCR18T2).
  Parts of the numerical calculations were performed using the facilities of the Supercomputer Center, the Institute for Solid State Physics, the University of Tokyo.
  The figures of the lattice structures were produced via VESTA package~\cite{Momma:db5098}.

%\nocite{*}

\providecommand{\noopsort}[1]{}\providecommand{\singleletter}[1]{#1}%

\clearpage
\includepdf[pages=1, link=true]{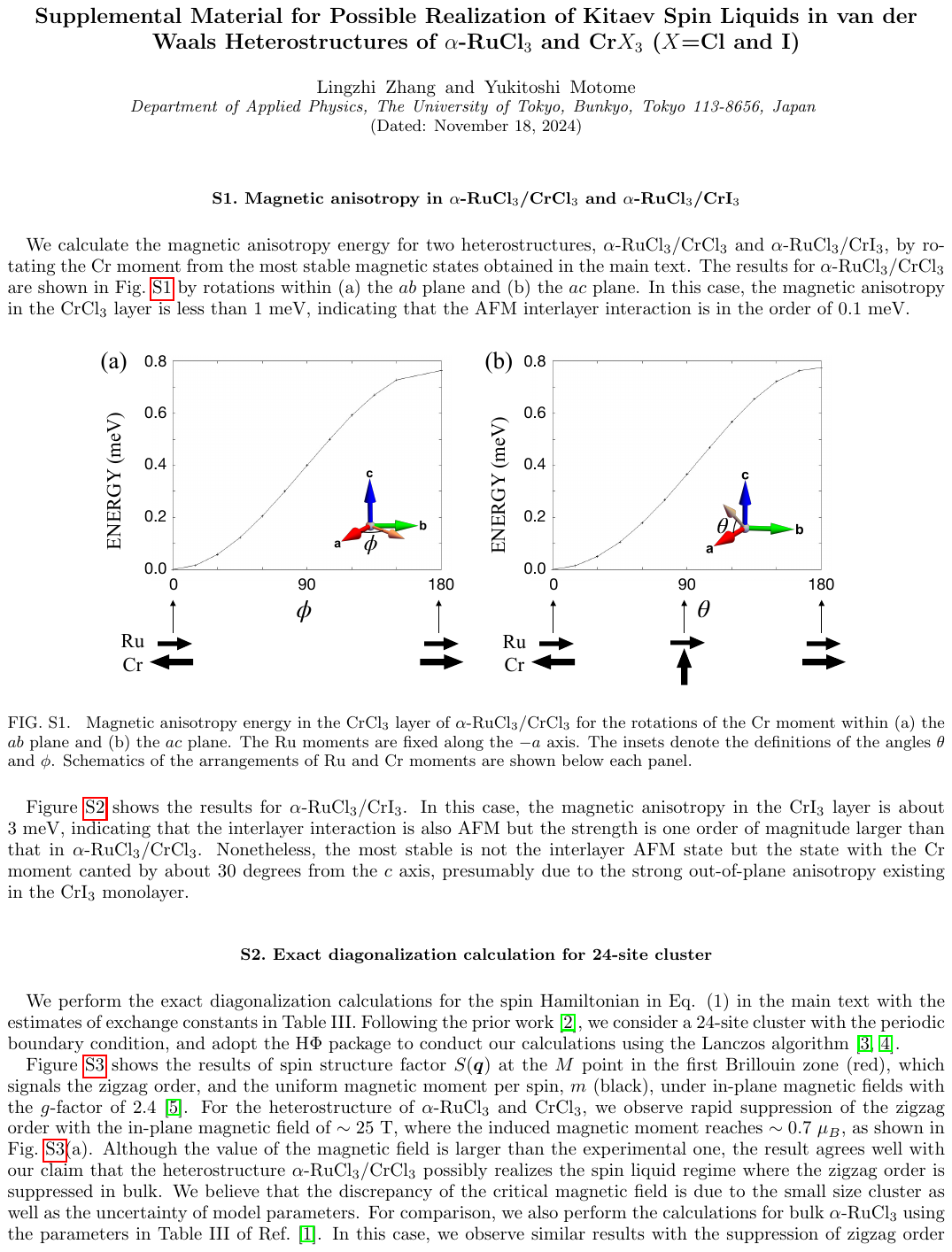}
\clearpage
\includepdf[pages=2, link=true]{SM.pdf}
\clearpage

\end{document}